\documentclass[10pt,preprint]{aastex}

\begin{document}

\title{Different Characteristics of the Bright Branches \\
of the Globular Clusters M3 and M13}

\author{Dong-Hwan Cho and Sang-Gak Lee}
\affil{Astronomy Program, SEES, Seoul National University,
Seoul 151-742, Korea}
\email{chodh@astro.snu.ac.kr, sanggak@astrosp.snu.ac.kr}
\and
\author{Young-Beom Jeon and Kyung Jin Sim}
\affil{Korea Astronomy Observatory, Daejeon 305-348, Korea}
\email{ybjeon@boao.re.kr}

\begin{abstract}
We carried out wide-field $B$$V$$I$ CCD photometric observations of the
globular clusters M3 (NGC 5272) and M13 (NGC 6205) using the Bohyun Optical
Astronomy Observatory
1.8-m telescope equipped with a SITe 2K CCD. We present color-magnitude
diagrams ($V$ vs. $B$$-$$V$, $V$ vs. $V$$-$$I$, and $V$ vs. $B$$-$$I$)
of M3 and M13. We have found asymptotic giant branch (AGB) bumps at
$V$$_{\rm AGB}^{\rm bump}$ = 14.85 ${\pm}$ 0.05 mag for M3 at
$V$$_{\rm AGB}^{\rm bump}$ = 14.25 ${\pm}$ 0.05 mag for M13. It is
found that AGB stars in M3 are more concentrated near the bump,
while those in M13 are scattered along the AGB sequence.
We identified the red giant branch (RGB) bump of M3
at $V$$_{\rm RGB}^{\rm bump}$ = 15.50 $\pm$ 0.05 mag and that of M13 at
$V$$_{\rm RGB}^{\rm bump}$ = 14.80 $\pm$ 0.05 mag through luminosity
functions and slope changes of the
integrated luminosity functions of M3 and M13. We have estimated the ratios
$R$ and $R_{\rm 2}$ for M3 and M13 and found that the value
of $R$ for M3 is larger than that for M13
while values of $R_{\rm 2}$ for M3 and M13 are similar and compatible with the
value expected from evolutionary theory when only normal horizontal
branch (HB) stars are used for estimation of $R$ and $R_{\rm 2}$ for M13.
However, we found that values of $R$ for M3 and M13 are similar
while the value of $R_{\rm 2}$ for M3 is larger than that for M13 when all
the HB stars are included for estimation of $R$ and $R_{\rm 2}$ for M13.
We have compared the observed RGB luminosity functions of M3 and M13 with the
theoretical RGB luminosity function of Bergbusch \& VandenBerg
at the same radial distances from the cluster centers
as used in the estimation of the ratios $R$ and $R_{\rm 2}$
for M3 and M13. We found ``extra stars'' belonging to M13
in the comparison of the observed RGB luminosity function of M13
and the theoretical RGB luminosity function of Bergbusch \& VandenBerg and
even in the comparison of the observed RGB luminosity functions of M3 and M13.
In the original definition of the ratio $R$ of Buzzoni
et al., $N_{\rm HB}$ corresponds to the lifetime of HB stars
in the RR Lyrae instability strip at log $T_{\rm eff}$ = 3.85. So, the smaller
$R$ value resulting for M13 compared with that for M3 in the case where only
normal HB stars are included in the estimation of $R$ and $R_{\rm 2}$ for 
M13 may be
partially caused by ``extra stars'', and the similar $R$ values for M3 and M13
in the case where the all HB stars are included in the estimation of
$R$ and $R_{\rm 2}$ for M13 may be caused by ``extra stars''
in the upper RGB of M13. If ``extra stars'' in the upper RGB of
M13 are caused by an effective ``deep mixing'' these facts
support the contention that an effective ``deep mixing'' could lead to
different HB morphologies between M3 and M13 and subsequent sequences.
\end{abstract}


\section{Introduction}

The globular clusters (GCs) M3 (NGC 5272) and M13 (NGC 6205) are a famous
second parameter GC
pair which have nearly the same metallicity $\langle$[Fe/H]$\rangle$
$\approx$ $-$1.49 $\pm$ 0.14 (e.g., Kraft et al. 1992; Sneden et al. 2004)
but possess different horizontal branch (HB) morphology. M3 has
both red HB (RHB) stars and blue HB (BHB) stars including RR Lyrae stars which
fill the instability region of the HB sequence (Sandage 1953; Buonanno et al.
1994; Ferraro et al. 1997a), while M13 has predominantly BHB stars which
extend below the
main sequence turnoff (MSTO) point. There are some gaps along the BHB star
sequence
and the cluster has some RR Lyrae stars but no RHB stars (Arp \& Johnson 1955;
Savedoff 1956; Paltrinieri et al. 1998).

The relative age difference between M3 and M13 was suggested as a second
parameter which controls the HB morphology difference between M3 and M13
(Rey et al. 2001),
although some researchers argue that the age difference between M3 and M13
is too small to
explain the HB morphology difference (Paltrinieri et al. 1998; Salaris \&
Weiss 2002). It is suggested by Sweigart (1997) that ``deep mixing'' could
cause the difference in HB morphologies if the ``deep mixing'' in two
GCs has occurred to different degrees. Cavallo \& Nagar (2000)
suggested that deep mixing,
which effectively occurred in M13 but not in M3, is a ``blue-tail second
parameter''
of M3 and M13, although Smith (2002) argued that deep mixing also
occurred in M3 using the merged data sets of the $\lambda$3883 CN band
strength of
red giant stars of M3. However Moehler et al. (2003) could not confirm
that the ``helium mixing'' scenario is a definite explanation for the origin
of the blue HB tail in M13. Moreover Caloi (2001) argued that there is no
evidence for a substantial increase in the surface helium content of HB
stars of M3 and M13, which contradicts the deep mixing and chemical
inhomogeneities of the red giant branch (RGB) stars of M13, and argued that
the most peculiar
giants may belong to the asymptotic giant branch (AGB) stars.
However, we should note that we cannot directly measure helium abundances
of late-type stars (RGB and AGB stars, main sequence stars, and normal
HB stars) of GCs
and that hot BHB stars in GCs show deficient helium abundances (Heber et al.
1986 and Moehler, Heber, \& Rupprecht 1997 for NGC 6752; Behr et al. 1999
for M13; Behr, Cohen, \& McCarthy 2000 for M15; Behr 2003 for M3, M13, M15,
M68, M92, and NGC 288) due to helium diffusion (Greenstein, Truran, \&
Cameron 1967; Michaud, Vauclair, \& Vauclair 1983).

AGB stars of GCs are evolved from HB stars and may reflect the different
characteristics of the HB stars of a given GC. However, the number of AGB stars
in a given GC is small due to the short lifetime of the AGB state and the
separation of AGB stars from RGB stars in a given GC was not easily
distinguished in previous photometric studies of GCs.
Therefore, little attention has been paid to the AGB stars of GCs so far.
We carried out wide-field $B$$V$$I$ CCD photometric observations
of M3 and M13 in order to obtain a sufficient number of AGB stars
accurately separated from RGB stars and to discover the different
characteristics of all bright branches in M3
and M13 which may reveal clues as to the second parameter of M3 and M13.
Here we present preliminary photometric results for the central 11${\farcm}$7
$\times$ 11${\farcm}$7 regions of M3 and M13.

In $\S$ 2 we present observations and data reduction for M3 and M13, and in
$\S$ 3 we present color-magnitude diagrams (CMDs) of these clusters. In $\S$ 4
we report identification of the RGB bumps of M3 and M13, and in $\S$ 5 we
present population ratios for the bright branches of M3 and M13. In $\S$ 6
we compare the observed RGB luminosity functions of M3 and M13 with
the theoretical RGB luminosity function of Bergbusch \& VandenBerg (2001),
and in $\S$ 7 we present a summary.

\section{Observations and Data Reduction}

$B$$V$$I$ CCD observations of the central 11${\farcm}$7 $\times$ 11${\farcm}$7
regions of M3 and M13 and three of Landolt's (1992) standard regions were
made using the Bohyun Optical Astronomy Observatory (BOAO) 1.8-m telescope
(f/8) during three continuous photometric nights between May 10 and
May 12, 2001. The detector attached to the Cassegrain focus of the telescope
was a SITe 2048 $\times$ 2048 CCD with a gain of 1.8 electrons ADU$^{{-}1}$
and readout noise of 7 electrons. The pixel size of the CCD is 24 $\micron$
square and the pixel scale is 0${\farcs}$344 pixel$^{{-}1}$ covering a
11${\farcm}$7 $\times$ 11${\farcm}$7 region of the sky.
The observational log of M3 and M13 is summarized in Table 1.

\placetable{tbl-1}

Preprocessing including bias subtraction scaled to overscan
regions, trimming of useless sections, flat fielding in each filter, and
cosmic ray rejection of CCD frames were carried out using an
IRAF/CCDRED package.

Standardization of photometric systems was made using three Landolt's
(1992) standard regions: SA 107 (centered near star 107-595), PG
1633$+$099, and PG 1323$-$086. Photometry of stars in the object frames of
M3 and M13 was carried out using the IRAF version of DAOPHOT
(Stetson, Davis, \& Crabtree 1990) via the point spread function fitting
method which is used for crowded field photometry.

Detailed standardization and reduction procedures will be published
in the next paper with results for $\sim$20$\arcmin$ $\times$ 20$\arcmin$
regions of M3 and M13.

\section{Color-Magnitude Diagrams}

\subsection{M3}

CMDs of M3 using color-magnitude data matched in all 3 bands are shown
in Figure 1. Owing to severe crowding, stars within a projected distance
from the cluster center $r$ ${<}$ 1${\farcm}$43 were excluded and so
the total number of stars plotted in the CMDs is 4834.

\begin{figure}
\figurenum{1} \epsscale{0.8} \plotone{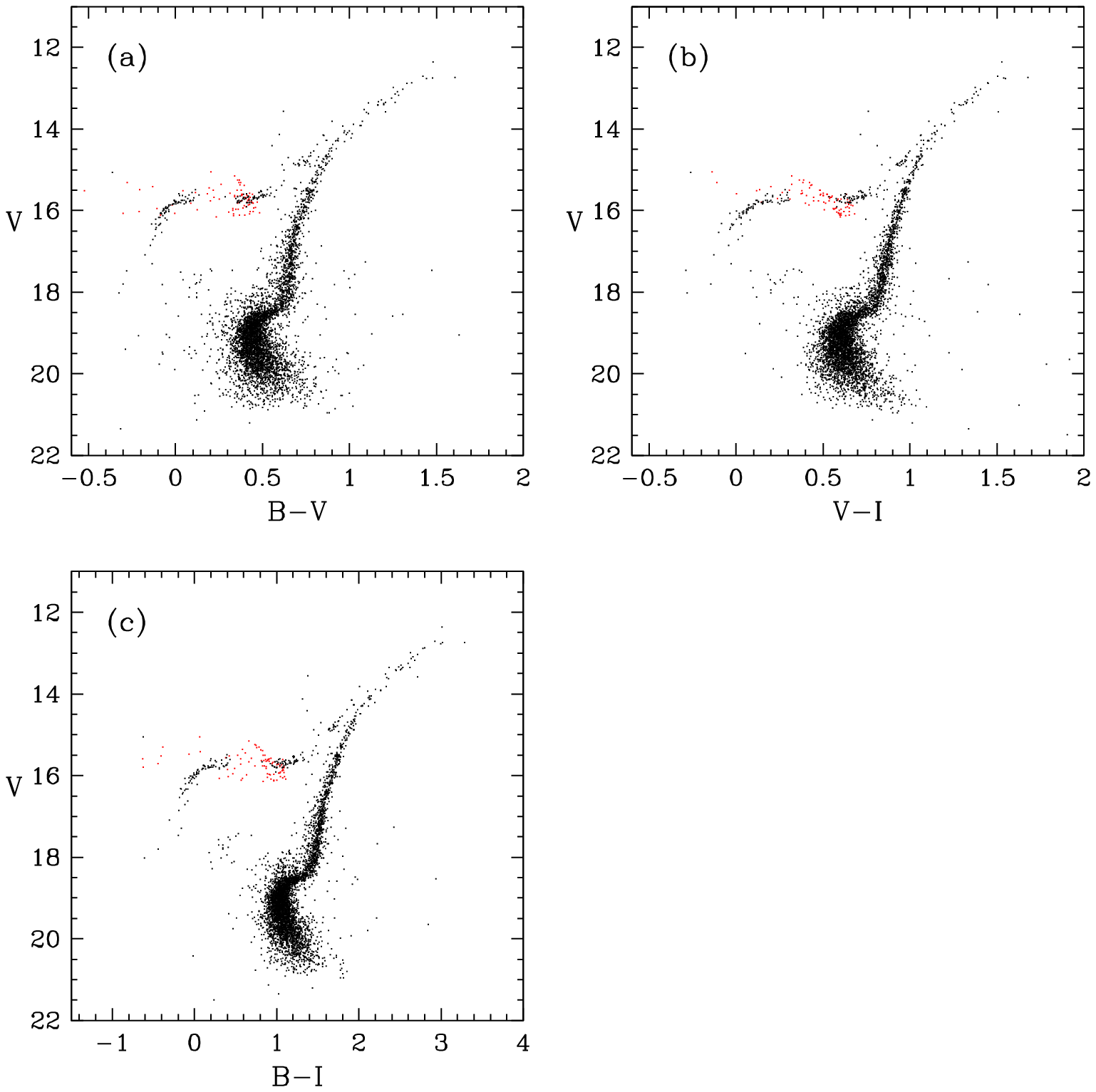}
\caption{Color-magnitude diagrams of M3. Only stars with projected distance
from the cluster center $r$ ${\ge}$ 1${\farcm}$43, whose total
number is 4834, are shown. Red points are 84 known RR Lyraes from the
observed field according to Clement et al.'s (2001) M3 variable stars
list. ({\it a}) $V$ vs. $B$$-$$V$ CMD. ({\it b}) $V$ vs. $V$$-$$I$ CMD.
({\it c}) $V$ vs. $B$$-$$I$ CMD.}
\label{fig1}
\end{figure}

\placefigure{fig1}

Since there is a possibility that the zero points in the magnitude
and color of M3 are not sufficiently accurate,
possibly due to fluctuating sky conditions
at BOAO, we examined the zero point offsets in the magnitude and color of M3 by
comparing mean fiducial sequences of our original CMDs and recent studies.
We then shifted the zero point offsets in magnitude and color from our original
CMD data. For the $V$ versus $B$$-$$V$ CMD, the comparison of mean fiducial
sequences
was made between our CMD and Rey et al.'s (2001) and for the $V$ versus
$V$$-$$I$ CMD, it was conducted between our CMD and Johnson \& Bolte's (1998).
According to these comparisons, the zero point offsets are ${\Delta}V$ =
0.15 $\pm$ 0.03 mag, $\Delta$($B$$-$$V$) = 0.155 $\pm$ 0.005 mag, and
$\Delta$($V$$-$$I$) = 0.075 $\pm$ 0.005 mag, where $\Delta$ is in the
sense of our study minus the other study. According to these comparisons
$\Delta$($B$$-$$I$) is 0.230 $\pm$ 0.007 mag. Thus the CMDs of M3 in Figure 1
are zero point-shifted CMDs according to the zero point offsets in
magnitude and color from the original CMDs of M3.

The internal photometric errors in magnitude and color have been estimated
by the photometric errors derived by DAOPHOT. According to these estimates
${\epsilon}(V)$, ${\epsilon}$($B$$-$$V$), ${\epsilon}$($V$$-$$I$), and
${\epsilon}$($B$$-$$I$) are ${\approx}$ 0.01 mag down to $V$ ${\approx}$
17.85 mag, and ${\epsilon}(V)$ and ${\epsilon}$($B$$-$$V$) are ${\approx}$
0.02 mag, ${\epsilon}$($V$$-$$I$) and ${\epsilon}$($B$$-$$I$) ${\approx}$
0.03 mag from $V$ ${\approx}$ 17.85 mag down to $V$ ${\approx}$ 19.35
mag, then ${\epsilon}(V)$ increases continuously up to ${\epsilon}(V)$
${\approx}$ 0.05 mag, ${\epsilon}$($B$$-$$V$), ${\epsilon}$($V$$-$$I$), and
${\epsilon}$($B$$-$$I$) increase to ${\approx}$ 0.08 mag from $V$ ${\approx}$
19.35 mag to $V$ ${\approx}$ 20.65 mag.

The characteristics of the CMDs of M3 presented in Figure 1 are as follows.
First, the separation between RGB and AGB stars is clearer and the
photometric accuracy of bright branch stars is greater than any other previous
photometric studies of M3. Second, the AGB and RGB sequences are clearly
separated best in the $V$ versus $B$$-$$I$ CMD among the three kinds of CMD
in Figure 1. AGB stars are concentrated near the AGB bump.
Third, stars are largely distributed along the RR Lyrae instability strip
and some stars are found below the level of the ZAHB (Zero Age Horizontal
Branch) in both the RHB and the BHB. When 84 known RR Lyraes (Clement et al. 2001)
are identified from the observed field, these latter stars are found to
be those in the CMDs in Figure 1 which are represented as red points. So,
these stars which are found below the level of ZAHB in both the RHB
and the BHB are RR Lyrae variable stars in their close minimum or minimum
pulsational stages in terms of their $V$ magnitude.

\subsection{M13}

The CMDs of M13 using color-magnitude data matched in all 3 bands are shown
in Figure 2. Owing to severe crowding, stars within a projected distance
from the cluster center $r$ ${<}$ 1${\farcm}$43 were excluded and so the
total number of stars plotted in the CMDs is 7056.

\begin{figure}
\figurenum{2} \epsscale{0.8} \plotone{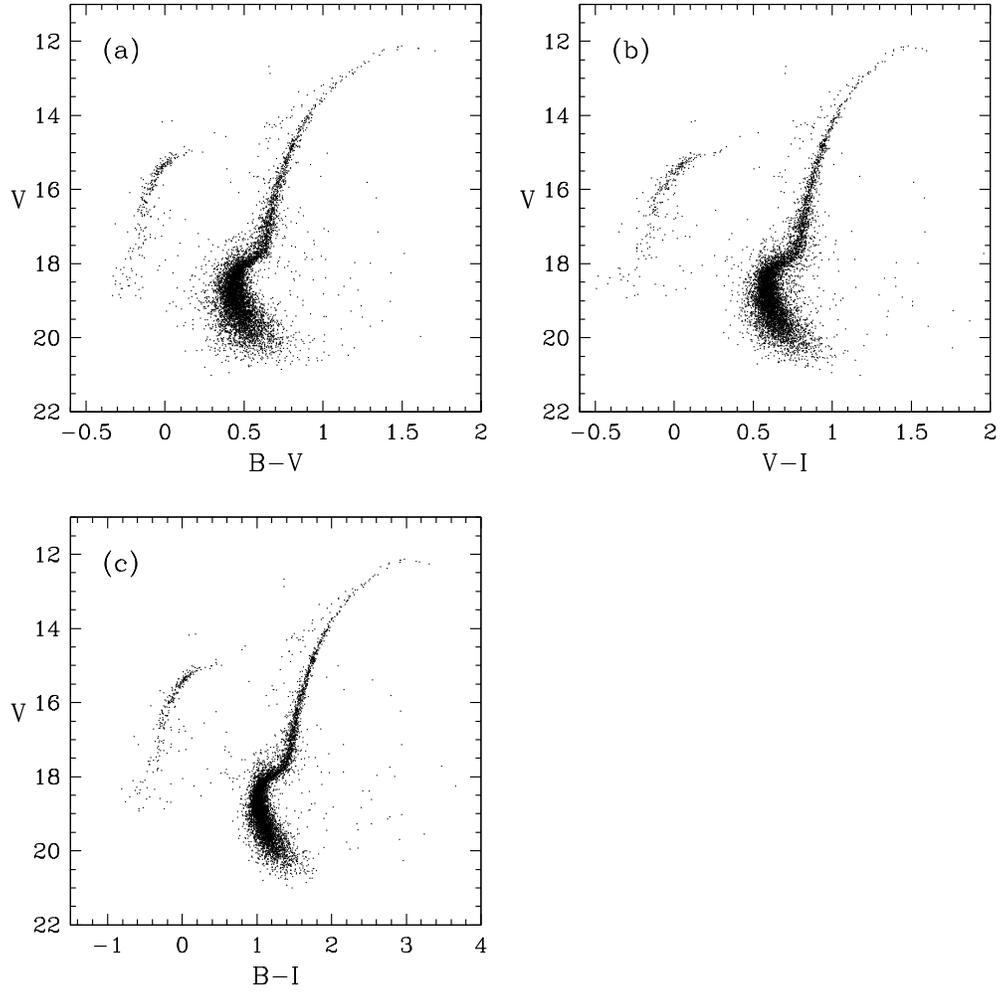}
\caption{Color-magnitude diagrams of M13. Only stars with projected distance
from the cluster center $r$ ${\ge}$ 1${\farcm}$43, whose total
number is 7056, are shown. ({\it a}) $V$ vs. $B$$-$$V$ CMD.
({\it b}) $V$ vs. $V$$-$$I$ CMD. ({\it c}) $V$ vs. $B$$-$$I$ CMD.}
\label{fig2}
\end{figure}

\placefigure{fig2}

As in the case of M3 there is a possibility that the
zero points in the magnitudes and colors of M13 are not accurate, possibly
due to the fluctuating sky conditions at BOAO, so we examined the zero point
offsets in the magnitude and color of M13. This was conducted by comparing
mean fiducial sequences of the original CMDs and recent studies and
shifting the zero point offsets in magnitude
and color from our original CMD data. For the $V$ versus $B$$-$$V$ CMD the
comparison of mean fiducial sequences was made between our CMD and Rey et al.'s
(2001) and for the $V$ versus $V$$-$$I$ CMD it was made between our CMD
and Johnson \& Bolte's (1998). According to these comparisons, the zero point
offsets are ${\Delta}V$ = $-$0.10 $\pm$ 0.03 mag, $\Delta$($B$$-$$V$) =
$-$0.060 $\pm$ 0.005 mag, and $\Delta$($V$$-$$I$) = 0.020 $\pm$ 0.005 mag,
where $\Delta$ is in the sense of our study minus the other study. 
Therefore, we have $\Delta$($B$$-$$I$) = $-$0.040 $\pm$ 0.007 mag.
Thus, the CMDs of M13 in Figure 2 are zero point-shifted CMDs according
to the above zero point offsets in magnitude and color from the original CMDs
of M13.

The internal photometric errors in magnitude and color have been estimated
by photometric errors derived by DAOPHOT. We have found
${\epsilon}(V)$, ${\epsilon}$($B$$-$$V$), ${\epsilon}$($V$$-$$I$), and
${\epsilon}$($B$$-$$I$) are ${\approx}$ 0.01 mag down to $V$ ${\approx}$
16.60 mag, and ${\epsilon}(V)$ and ${\epsilon}$($B$$-$$V$) are ${\approx}$
0.02 mag, ${\epsilon}$($V$$-$$I$) and ${\epsilon}$($B$$-$$I$) ${\approx}$
0.03 mag from $V$ ${\approx}$ 16.60 mag down to $V$ ${\approx}$ 18.10
mag, then ${\epsilon}(V)$ increases continuously up to ${\epsilon}(V)$
${\approx}$ 0.07 mag, ${\epsilon}$($B$$-$$V$), ${\epsilon}$($V$$-$$I$), and
${\epsilon}$($B$$-$$I$) increase to ${\approx}$ 0.10 mag from $V$ ${\approx}$
18.10 mag to $V$ ${\approx}$ 20.35 mag.

The characteristics of the CMDs of M13 presented in Figure 2 are as follows.
First, the BHB, RGB, AGB, and main sequences are well populated from the RGB
tip to about 2.0 mag below the MSTO. Second, the RGB and AGB stars are
clearly separated and they are distinguished best in the $V$ versus $B$$-$$I$
CMD. AGB stars are scattered along the AGB sequence. Third, a few stars at
the RGB tip tend slightly to the redder and fainter direction.
Last, BHB stars are well populated down to the MSTO and there are at least
two gaps along the BHB sequence.

\section{The RGB Bump}

We examined the existence of RGB bumps in M3 and M13, and identified
the location of RGB bumps in M3 and M13 through the standard method described
in Fusi Pecci et al. (1990) from $V$ versus $B$$-$$V$ CMDs of
Figure 1{\it a} in the case of M3 and Figure 2{\it a} in the
case of M13.

The method for detecting RGB bumps of M3 and M13 is as follows. From
the $V$ versus $B$$-$$V$ CMDs of M3 and M13 we first rejected clear HB
and AGB stars to accelerate the procedure of rejection of
non-RGB stars from the RGB sequences. Then, we divided the RGB stars of M3
and M13 into bins of size 0.25 mag. In order to reject outlying
field stars we calculated the mean and standard deviation of each bin
and rejected stars lying outside of 2.5~${\sigma}$ of the mean value
of each bin. We recalculated the mean and standard deviation of each
bin for which the outlying field stars were rejected according to the
2.5~${\sigma}$ rejection criterion, and rejected further stars
lying outside of the new 2.5~${\sigma}$ of the new mean value of each
bin and iterated these procedures several times until the mean and
standard deviation of each bin converged.
After rejecting the outlying field stars along the RGB sequences of M3 and
M13, we constructed integrated luminosity functions and differential
luminosity functions of RGB sequences in bins of size 0.05 mag.
From these luminosity functions we identified the RGB bump where the peak
in the differential luminosity function and the slope change in the
integrated luminosity function coincide.

The integrated luminosity function and the differential luminosity
function of M3 are shown in Figure 3. Figure 3 shows that the position
of the RGB bump of M3 is $V$$_{\rm RGB}^{\rm bump}$ = 15.50 ${\pm}$ 0.05 mag,
which is in good agreement with the value 15.45 ${\pm}$ 0.05 mag of Ferraro
et al (1999).
The integrated luminosity function and the differential luminosity
function of M13 are shown in Figure 4. They show that the
position of the RGB bump of M13 is $V$$_{\rm RGB}^{\rm bump}$ = 14.80
${\pm}$ 0.05 mag. This bump position is slightly fainter than the 14.70 ${\pm}$
0.05 mag determination of Bono et al. (2001) but is in good agreement with
the 14.75 ${\pm}$ 0.07 mag value of Ferraro et al. (1999). In Figures 3 and 4
arrows indicate RGB bump
positions and the dotted lines in the integrated luminosity functions
indicate slope changes above and below the RGB bump positions.

\begin{figure}
\figurenum{3} \epsscale{0.5} \plotone{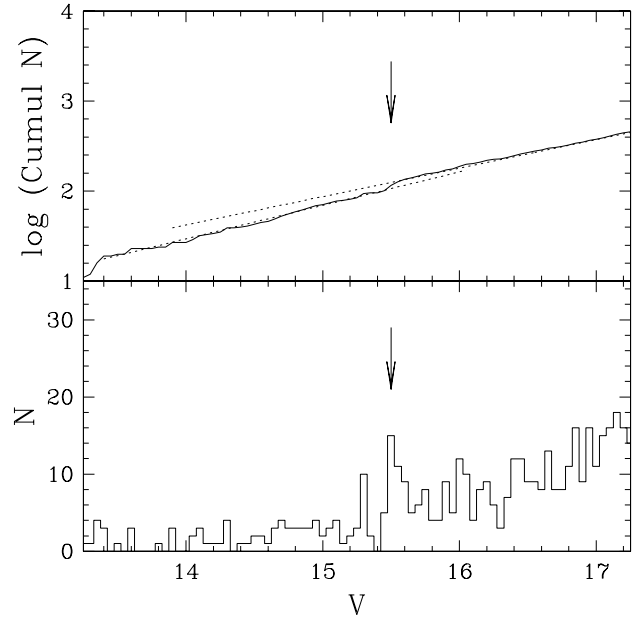} 
\caption{Integrated luminosity function and differential
luminosity function of M3. Arrows in each panel indicate RGB bump
position and dotted lines in the upper panel indicate slope change
of the integrated luminosity function above and below the RGB bump
position.}
\label{fig3}
\end{figure}

\begin{figure}
\figurenum{4} \epsscale{0.5} \plotone{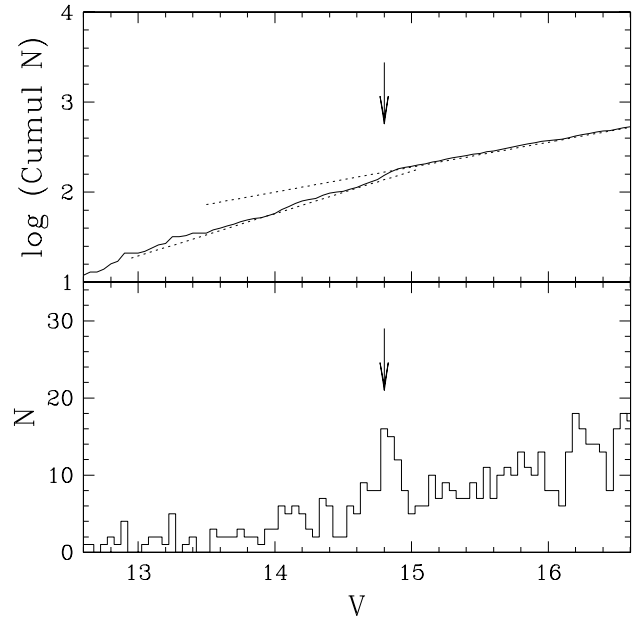}
\caption{Same as Fig.~3 but for M13.}
\label{fig4}
\end{figure}

\placefigure{fig3}
\placefigure{fig4}

\section{Population Ratios}

We calculated the population ratios $R$ (= $N_{\rm HB}$/$N_{\rm RGB}$),
and $R_{\rm 2}$ (= $N_{\rm AGB}$/$N_{\rm HB}$) for M3 and M13 using their
$V$ versus $B$$-$$I$ CMDs in Figures 1{\it c} and 2{\it c} for M3 and M13,
respectively, in which AGB and RGB sequences are the most clearly separated
from each other among the three kinds of CMD. Here, $N_{\rm HB}$ is the
number of normal HB stars including BHB stars, RR Lyrae variable stars,
and RHB stars, and $N_{\rm RGB}$ is the number
of RGB stars brighter than the mean luminosity level of HB stars
(= ${\langle}V_{\rm HB}{\rangle}$), and $N_{\rm AGB}$ is the number of
AGB stars which are brighter than the AGB bump position.
In order to obtain population ratios $R$ and $R_{\rm 2}$ from complete
samples of M3 and M13, we removed the stars with a projected distance
from the cluster center $r$ smaller than given values. In the case of M3
stars, they are complete down to $V$ = 17.65 mag when $r$ $\ge$ 2${\farcm}$6
and complete down to $V$ = 19.00 mag when $r$ $\ge$ 3${\farcm}$2
due to severe crowding in the cluster center region.
Since in M3 $V$ = 17.65 mag is fainter than the normal HB level
($V$ = 16.75 mag) and the RGB magnitude limit, which includes all the RGB stars
used in the $R$ ratio ($V$ ${<}$ 16.01 $\pm$ 0.07 mag) and the RGB luminosity
function ($V$ ${<}$ 17.53 mag) in $\S$ 6, we removed stars with $r$
${<}$ 2${\farcm}$6 in the derivation of the population ratios $R$ and
$R_{\rm 2}$
for M3. According to Harris (1996), the half-mass radius ($r_{h}$) of M3 is
1${\farcm}$12 and $r$ = 2${\farcm}$6 corresponding to
$\sim$2${\farcm}$32$r_{h}$.
In the case of M13, stars are complete down to $V$ = 16.00 mag (luminosity
level of the first gap in the BHB of M13) when $r$ $\ge$ 2${\farcm}$5 and
complete down to $V$ = 19.00 mag when $r$ $\ge$ 3${\farcm}$2 due to severe
crowding in the cluster center region. Since in M13 hot
BHB stars extend down to $V$ $\approx$ 19.00 mag and we also
consider the population ratios $R$ and $R_{\rm 2}$, which include all
the HB stars of M13, we removed stars with $r$ ${<}$ 3${\farcm}$2 in
the derivation of the population ratios $R$ and $R_{\rm 2}$ of M13.
According to Harris (1996), $r_{h}$ of M13 is 1${\farcm}$49 and $r$ =
3${\farcm}$2 corresponds to $\sim$2${\farcm}$15$r_{h}$.

The AGB bump positions of
M3 and M13 are indicated by horizontal arrows in Figures 5 and 6,
respectively, and the positions are $V$$_{\rm AGB}^{\rm bump}$ = 14.85 ${\pm}$
0.05 mag for M3 and $V$$_{\rm AGB}^{\rm bump}$ = 14.25 ${\pm}$ 0.05
mag for M13. However, AGB stars in M3 are more concentrated near the
bump and show a slightly slanted AGB sequence above RGB, while AGB
stars in M13 are less concentrated near the bump and show a scattered
AGB sequence above RGB. These AGB features of M3 and M13 are
consistent with the theoretical predictions by Castellani,
Chieffi, \& Pulone (1991). According to their Figure 9 which
shows the comparison between theoretical evolutionary tracks of
HB and AGB stars for $Z$ = 0.001 and the CMD of M5 whose metallicity
([Fe/H] = $-$1.13) is known to be close to the metallicities
of M3 and M13, HB stars whose mass range is appropriate for the
HB stars of M3 produce a redder and tighter AGB sequence than
those HB stars whose mass range is appropriate for the HB stars of
M13, which create a bluer and relatively wider AGB sequence.
AGB stars brighter than AGB bumps are enclosed by small open
circles in Figures 5 and 6 for M3 and M13, respectively.
According to Figure 5 in the CMD of M3, AGB stars are separated from the
RGB starting from the AGB bump position up to $\sim$0.5 mag below
the RGB tip of M3, and according to
Figure 6 in CMD of M13, AGB stars are separated from the RGB starting
from the AGB bump position up to $\sim$1.0 mag below the RGB tip of M13.
We have found that the total number of AGB stars of M3 is 30 and that of M13
is 24. So, $N_{\rm AGB}$ of M3 is 15 in the $r$ $\ge$ 2${\farcm}$6 range
and $N_{\rm AGB}$ of M13 is 9 in the $r$ $\ge$ 3${\farcm}$2 range,
respectively.

\begin{figure} 
\figurenum{5} \epsscale{0.8} \plotone{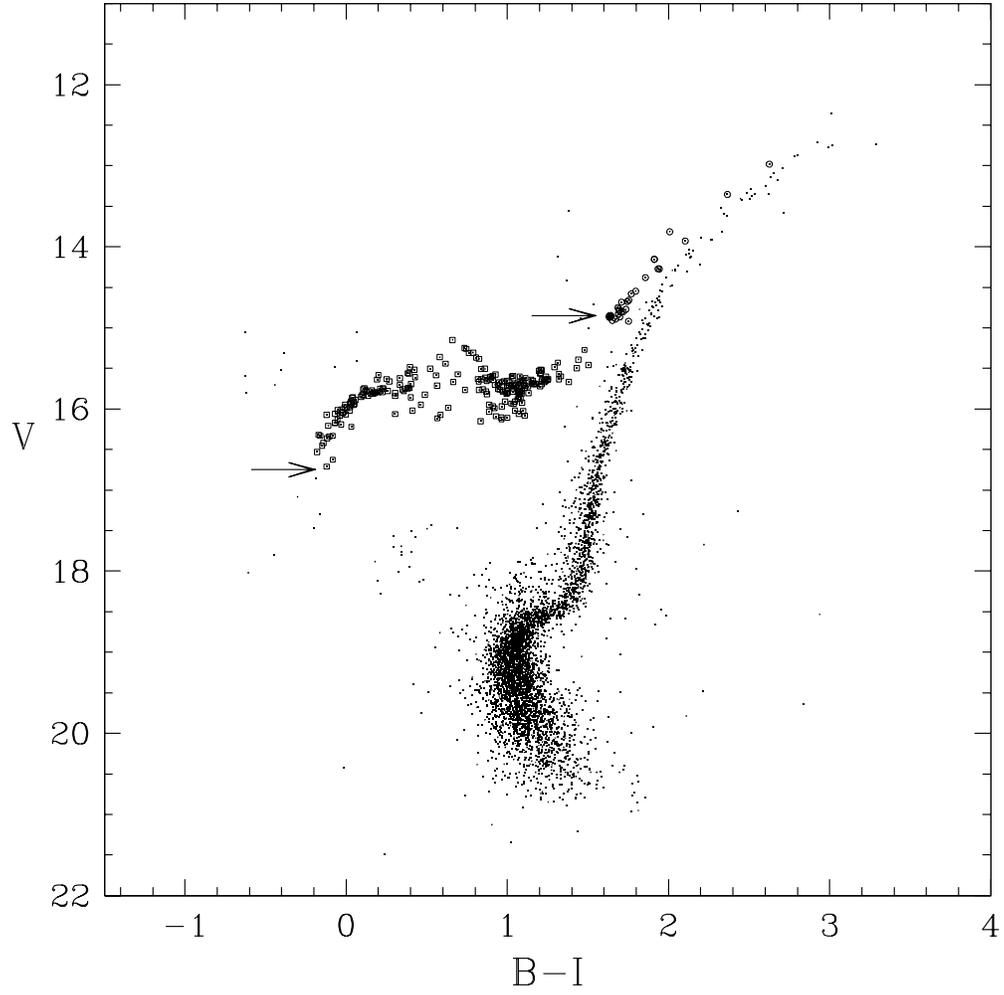}
\caption{$V$ vs. $B$$-$$I$ CMD of M3 denoting normal HB stars and
AGB stars brighter than the AGB bump position. Normal HB stars are enclosed
by small open squares and AGB stars brighter than the AGB bump position
are enclosed by small open circles. Horizontal arrow at the left side of
the BHB indicates lower limit of normal HB stars and horizontal arrow at the
base of the AGB indicates AGB bump position.}
\label{fig5}
\end{figure}

\begin{figure}
\figurenum{6} \epsscale{0.8} \plotone{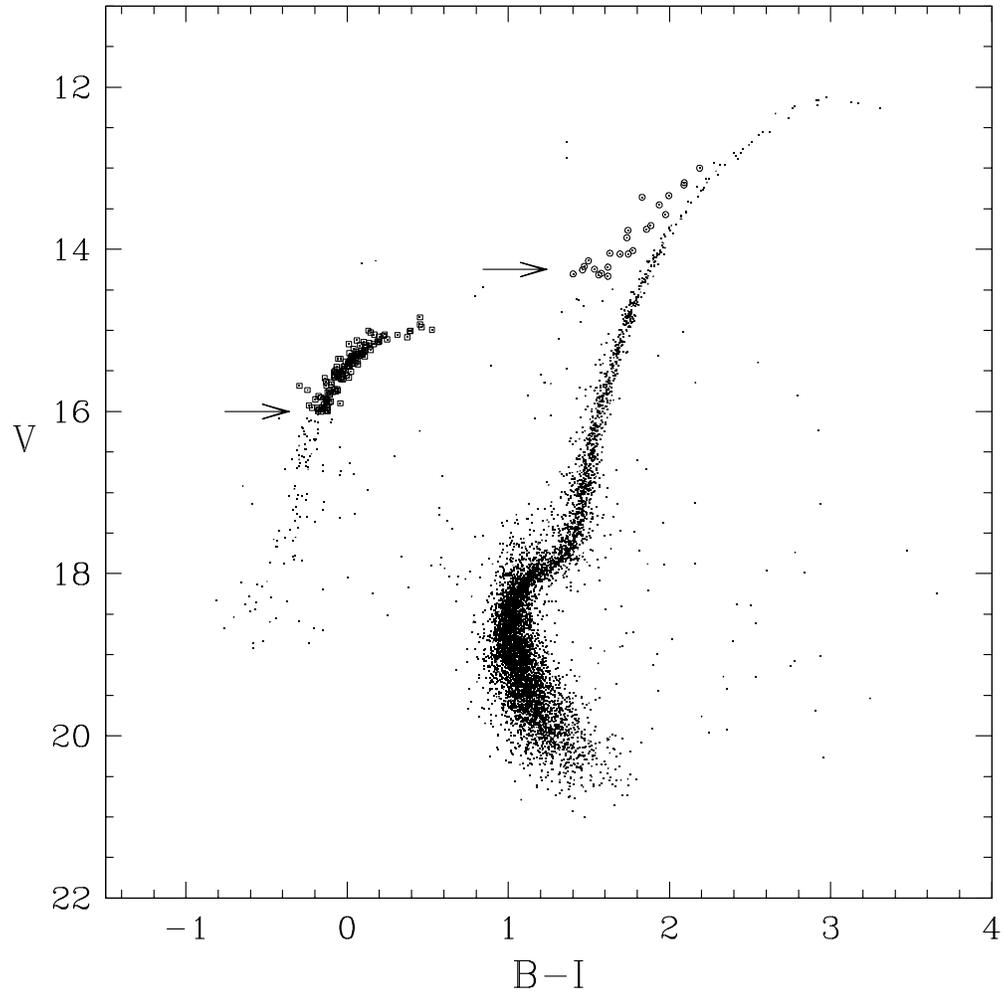}
\caption{Same as Fig.~5 but, for M13. Symbols and arrows are the same as
in Fig.~5.}
\label{fig6}
\end{figure}

\placefigure{fig5}
\placefigure{fig6}

In order to calculate population ratios we must know
${\langle}V_{\rm HB}{\rangle}$
(mean luminosity level of HB), ${\Delta}BC$ (differential bolometric
correction between ${\langle}V_{\rm HB}{\rangle}$ at
log $T_{\rm eff}$ = 3.85 and
RGB at the ${\langle}V_{\rm HB}{\rangle}$ level). Since the metallicities of
M3 and M13 are nearly the same, we assumed equal ${\Delta}BC$ for M3 and M13.
We took two kinds of value of ${\Delta}BC$ from other
studies. One is ${\Delta}BC$ = 0.11 mag from Ferraro et al. (1997a)
and the other is ${\Delta}BC$ = 0.29 mag from Sandquist (2000).
Ferraro et al. (1997a) adopted ${\Delta}BC$ = 0.11 mag from private
communication with O. Straniero. Sandquist (2000) adopted ${\Delta}BC$
= 0.29 mag from the fitting formula:

\begin{equation}
{\Delta}BC = 0.709 + 0.548[{\rm M/H}] + 0.229[{\rm M/H}]^{2} +
0.034[{\rm M/H}]^{3}~ ,
\end{equation}

where [M/H] is the $\alpha$-element enhanced global metallicity of GCs and
he adopted a constant value [$\alpha$/Fe] = +0.3 for $\alpha$-element
enhancement of GCs. Sandquist (2000) derived the above fitting formula using
the HB models of Dorman (1992) in conjunction with the isochrones of
Bergbusch \& VandenBerg (1992).

In the case of ${\langle}V_{\rm HB}{\rangle}$, we calculated
${\langle}V_{\rm HB}{\rangle}$
of M3 first and derived ${\langle}V_{\rm HB}{\rangle}$ of M13
by matching the AGB bumps and RGB bumps of M3 and M13 because M13
has no RHB stars and is known to have only nine RR Lyrae
variable stars (Preston, Shectman, \& Beers 1991;
Kopacki, Ko{\l}aczkowski, \& Pigulski 2003). For M3,
${\langle}V_{\rm HB}{\rangle}$ was calculated from 22 stars from the BHB edge
region [0.005 ${\leq}$ ($B$$-$$V$) ${\leq}$ 0.100 and 15.475 ${\leq}$ $V$
${\leq}$ 15.880] and 34 stars from the RHB edge region [0.345 ${\leq}$
($B$$-$$V$) ${\leq}$ 0.465 and 15.475 ${\leq}$ $V$ ${\leq}$ 15.850] excluding
RR Lyrae variable stars.
${\langle}V_{\rm HB}{\rangle}$ in the BHB edge region is
${\langle}V_{\rm HB}{\rangle}$$_{\rm BE}$
= 15.732 ${\pm}$ 0.080 mag and that in the RHB edge region is
${\langle}V_{\rm HB}{\rangle}$$_{\rm RE}$ = 15.716 ${\pm}$ 0.063 mag and the
total average is ${\langle}V_{\rm HB}{\rangle}$ = 15.72 ${\pm}$ 0.07 mag.
We adopted ${\Delta}V$ = 0.70 mag to derive the mean
HB level of M13 assuming the same intrinsic mean HB levels in both
clusters, for which the magnitude difference is found
between the RGB bumps of M3 and M13 in this study. Then the BHBs of M3 and
M13 coincide nicely. Therefore,
${\langle}V_{\rm HB}{\rangle}$ of M13 was derived to be
${\langle}V_{\rm HB}{\rangle}$ = 15.02 ${\pm}$ 0.10 mag.
However, if we adopted ${\Delta}V$ = 0.60 mag from the magnitude difference
between the AGB bumps of M3 and M13, ${\langle}V_{\rm HB}{\rangle}$ of M13 is
derived to be ${\langle}V_{\rm HB}{\rangle}$ = 15.12 ${\pm}$ 0.10 mag.
However, in this case, the actual HB of M13 was $\sim$0.10 mag brighter
than that of M3 and the BHBs of M3 and M13 do not coincide.
Since the magnitude of the AGB bump position is nearly constant
independent of metallicity and helium abundance
(Castellani et al. 1991; Bono et al. 1995), this point may indicate
that ${\langle}V_{\rm HB}{\rangle}$ of M13 is $\sim$0.10 mag brighter than
that of M3. So, we finally adopted ${\langle}V_{\rm HB}{\rangle}$ of M13 as
${\langle}V_{\rm HB}{\rangle}$ = 15.02 ${\pm}$ 0.10 mag.

According to Cassisi et al. (2001) the difference between the bottom
luminosity of the AGB clump and the luminosity of the ZAHB level
($\Delta$$M_{V}$(AGB-HB))
appears largely independent of variations in the assumed
progenitor mass and/or original helium content ($Y$). But, because they did
not present the exact quantitative dependency of $\Delta$$M_{V}$(AGB-HB)
on the progenitor mass and original helium content we cannot know how
much $\Delta$$M_{V}$(AGB-HB) is independent of variations in the
assumed progenitor mass and/or original helium content. According to
Cassisi \& Salaris (1997) at a given $\alpha$-element enhanced global
metallicity [M/H] the difference in visual
magnitude between the RGB bump and the ZAHB level
($\Delta$$V$$_{\rm HB}^{\rm bump}$ = ($V$$_{\rm RGB}^{\rm bump}$ $-$
$V_{\rm ZAHB}$))
is weakly dependent on the original helium content and the age of the stellar
system from their exact quantitative argumentations and quite negligible
from the range of the expected efficiency of the mass-loss phenomenon
in real RGB stars. And they argue that $\Delta$$V$$_{\rm HB}^{\rm bump}$
can be used as a tentative guess to estimate the HB luminosity level
in those GCs in which the HB morphology is poorly populated in the RR
Lyrae instability strip or in which the HB morphology is quite blue or
quite red. Also, according to Cassisi, Degl'Innocenti, \& Salaris (1997)
$\Delta$$V$$_{\rm HB}^{\rm bump}$ is weakly affected by the helium and
heavy-element diffusion (the maximum difference is $\approx$0.06 mag).
So, at this stage it seems more reliable to use the magnitude difference
between the RGB bumps of M3 and M13 in deriving
${\langle}V_{\rm HB}{\rangle}$ of M13 than the magnitude difference between
the AGB bumps of M3 and M13.

From ${\langle}V_{\rm HB}{\rangle}$ and ${\Delta}BC$ of M3 and M13 we counted
the numbers of RGB stars brighter than the HB levels. In the case of M3, since
${\langle}V_{\rm HB}{\rangle}$ = 15.72 ${\pm}$ 0.07 mag,
$N_{\rm RGB}$ = 87$_{\rm -5}^{\rm +1}$ when ${\Delta}BC$ = 0.11 mag
($V$ ${<}$ 15.83 ${\pm}$ 0.07 mag) and $N_{\rm RGB}$ = 96$_{\rm -4}^{\rm +10}$
when ${\Delta}BC$ = 0.29 mag ($V$  ${<}$ 16.01 ${\pm}$ 0.07 mag) in the
$r$ $\ge$ 2${\farcm}$6 range. In the case of M13, since
${\langle}V_{\rm HB}{\rangle}$ = 15.02 ${\pm}$ 0.10 mag,
$N_{\rm RGB}$ = 74$_{\rm -5}^{\rm +10}$ when ${\Delta}BC$ = 0.11 mag
($V$ ${<}$ 15.13 ${\pm}$ 0.10 mag) and $N_{\rm RGB}$ = 92$_{\rm -11}^{\rm +4}$
when ${\Delta}BC$ = 0.29 mag ($V$ ${<}$ 15.31 ${\pm}$ 0.10 mag) in the
$r$ $\ge$ 3${\farcm}$2 range.
Here, errors in $N_{\rm RGB}$ arise from the magnitude
errors of ${\langle}V_{\rm HB}{\rangle}$ of M3 and M13.

In Figures 5 and 6 the horizontal arrows to the left of the BHBs of
M3 and M13 indicate the lower limits of normal HB levels of M3 and M13.
In the case of M3 this is $V$ = 16.75 mag and is roughly 1 mag below the
${\langle}V_{\rm HB}{\rangle}$ of M3. In the case of M13 this is $V$
= 16.00 mag, roughly 1 mag below the ${\langle}V_{\rm HB}{\rangle}$
of M13 and roughly the same position as the first gap in the BHB of M13
reported by Grundahl, VandenBerg, \& Andersen (1998) and Ferraro et al.
(1997b). Normal HB stars of
M3 and M13 are enclosed by small open squares in Figures 5 and 6.
According to the original definition of $R$ (= $N_{\rm HB}$/$N_{\rm RGB}$)
in Buzzoni et al. (1983), the
HB lifetime is the time spent by the HB stars in the RR Lyrae instability
strip at log $T_{\rm eff}$ = 3.85. Therefore, if all the HB stars (RHB stars,
RR Lyrae stars, BHB stars, and extended BHB stars) have the same lifetimes
as the HB stars in the RR Lyrae instability strip at log $T_{\rm eff}$
= 3.85 one can include all the HB stars in $N_{\rm HB}$ (to estimate the
HB lifetime at log $T_{\rm eff}$ = 3.85) to derive $R$ and the helium
abundance of a given GC. However, it later turned out that hot HB stars
of Galactic GCs have a more prolonged lifetime than the HB stars in the
RR Lyrae instability strip at log $T_{\rm eff}$ = 3.85, which is almost
identical to that of the HB stars redder than the RR Lyrae instability
strip (Castellani et al. 1994; Zoccali et al. 2000). So, one must correct
the prolonged
lifetime of hot HB stars of individual Galactic GCs which have BHB stars
and extended BHB stars with or without RHB stars in the derivation of
$R$ and the helium abundance of individual Galactic GCs. According to the
theoretical hot HB models of
Castellani et al. (1994), the lifetime of hot HB stars increases up to
$\sim$30$\%$ with respect to the HB stars in the RR Lyrae instability
strip and, according to the HB models of Zoccali et al. (2000), the
mean lifetime of blue HB stars is about 20$\%$ larger than that
of the HB stars in the RR Lyrae instability strip, which is almost
identical to that of the HB stars redder than the RR Lyrae instability strip.
Moreover, according to Zoccali et al. (2000), the $R$ values of Galactic GCs
with blue HB morphologies are expected to be $\approx$0.25 units higher
than those with RHBs.

However, Buzzoni et al. (1983), Caputo, Martinez Roger, \& Paez (1987), and
Sandquist (2000) did not know the prolonged lifetime of hot HB stars and
they did not take account of the prolonged lifetime of hot HB stars in
the derivation of helium abundances of Galactic GCs from the population
ratio $R$. Therefore, they included all the HB stars including hot HB
stars in $N_{\rm HB}$ and did not correct the prolonged lifetime of hot
HB stars by reducing $R$ (or $N_{\rm HB}$) by some factor related to
the prolonged lifetime of hot HB stars when the GCs have hot HB stars, and
just averaged out the all the values of helium abundances of Galactic
GCs to obtain the mean helium abundance of Galactic GCs.
This fact was recently noticed by Zoccali et al. (2000), Cassisi, Salaris,
\& Irwin (2003), and Salaris et al. (2004) in the derivation of $R$
ratios or helium abundances of Galactic GCs. So, Cassisi et al. (2003)
corrected the prolonged lifetime of hot HB stars when the GCs have hot HB
stars in the derivation of helium abundances of Galactic GCs using the
data of Sandquist (2000) and Zoccali et al. (2000). But, because the number
of GCs having hot HB stars is not so great, this correction of the
prolonged lifetime of hot HB stars did not greatly affect the mean
helium abundance of Galactic GCs. In the case of Salaris et al. (2004)
because they did not correct the prolonged lifetime of hot HB stars in
the derivation of helium abundances of Galactic GCs, they got a large
scatter of helium abundances of metal-poor GCs which in many cases have
hot HB stars.

So, in one method of correction for the prolonged lifetime of the hot HB stars
of M3 and M13 we marked only the normal HB stars in Figures 5 and 6 and
included only those stars in the first calculations of the $R$ and
$R_{\rm 2}$ ratios for M3 and M13.
However, if we include all the HB stars in the calculations of
$R$ and $R_{\rm 2}$ we must consider the prolonged lifetime of
extended BHB stars and reduce $N_{\rm HB}$ in accordance if we want to 
derive the true helium abundance and correct value of $R_{\rm 2}$ for a 
given GC.
According to Figure 5 the total number of normal HB stars of M3 is 215
and according to Figure 6 the total number of normal HB stars for M13 is
138. If we include all the HB stars of M13 the total number of HB stars for
M13 is 247. However, $N_{\rm HB}$ for M3 is 123 in the $r$ $\ge$ 2${\farcm}$6
range and $N_{\rm HB}$ for M13 is 57 in the $r$ $\ge$ 3${\farcm}$2
range.
In the case of M13, if we include all the HB stars
in the population ratios, $N_{\rm HB}$ for M13 is 127 in the
$r$ $\ge$ 3${\farcm}$2 range.

So the population ratios of M3 and M13 are as follows. For M3 $R$ =
$N_{\rm HB}$/$N_{\rm RGB}$ is = 123/(87$_{\rm -5}^{\rm +1}$) =
1.414$_{\rm -0.016}^{\rm +0.086}$ if ${\Delta}BC$ = 0.11 mag is adopted,
$R$ = $N_{\rm HB}$/$N_{\rm RGB}$ is = 123/(96$_{\rm -4}^{\rm +10}$)
= 1.281$_{\rm -0.121}^{\rm +0.056}$ if ${\Delta}BC$ = 0.29 mag is adopted,
and $R_{\rm 2}$ = $N_{\rm AGB}$/$N_{\rm HB}$ is = 15/123 = 0.122 in
the $r$ $\ge$ 2${\farcm}$6 range. Here, errors
arise from the ${\langle}V_{\rm HB}{\rangle}$ error of M3. These results are
briefly summarized in Table 2.

\placetable{tbl-2}

For M13 when we include only normal HB stars in $N_{\rm HB}$,
$R$ = $N_{\rm HB}$/$N_{\rm RGB}$ is = 57/(74$_{\rm -5}^{\rm +10}$)
= 0.770$_{\rm -0.091}^{\rm +0.056}$ if ${\Delta}BC$ = 0.11 mag is adopted,
 $R$ = $N_{\rm HB}$/$N_{\rm RGB}$ is = 57/(92$_{\rm -11}^{\rm +4}$)
= 0.620$_{\rm -0.026}^{\rm +0.084}$ if ${\Delta}BC$ = 0.29 mag is adopted,
and $R_{\rm 2}$ = $N_{\rm AGB}$/$N_{\rm HB}$ is = 9/57 = 0.158 in the
$r$ $\ge$ 3${\farcm}$2 range. Here,
errors arise from the ${\langle}V_{\rm HB}{\rangle}$ error of M13. These 
results are briefly summarized in Table 3.

\placetable{tbl-3}

For M13 when we include all the HB stars in $N_{\rm HB}$,
$R$ = $N_{\rm HB}$/$N_{\rm RGB}$ is = 127/(74$_{\rm -5}^{\rm +10}$)
= 1.716$_{\rm -0.204}^{\rm +0.125}$ if ${\Delta}BC$ = 0.11 mag is adopted,
 $R$ = $N_{\rm HB}$/$N_{\rm RGB}$ is = 127/(92$_{\rm -11}^{\rm +4}$)
= 1.380$_{\rm -0.057}^{\rm +0.188}$ if ${\Delta}BC$ = 0.29 mag is adopted,
and $R_{\rm 2}$ = $N_{\rm AGB}$/$N_{\rm HB}$ is = 9/127 = 0.071 in the
$r$ $\ge$ 3${\farcm}$2 range. Here,
errors arise from the ${\langle}V_{\rm HB}{\rangle}$ error of M13. These 
results are briefly summarized in Table 4.

\placetable{tbl-4}

\section{The RGB Luminosity Function}

We compared the RGB luminosity functions of M3 and M13 with the theoretical
RGB luminosity function of Bergbusch \& VandenBerg (2001) in order to
determine whether there are ``extra stars'' in the RGB luminosity functions of
M3 and M13. The RGB luminosity function data were taken from those used
in $\S$ 4 to examine the existence of RGB bumps in M3 and M13 and only
the bin size was changed to 0.25 mag to reduce the small number statistical
effects in the brighter regions where the number of stars is small.
As in the case of the population ratios for M3 and M13, in order to obtain the
observed RGB luminosity functions of M3 and M13 from the same complete
samples used in the population ratios and to directly compare them with
the population ratio $R$ for M3 and M13
we only used stars for which $r$ $\ge$ 2${\farcm}$6 from the cluster center in 
the case of M3 and $r$ $\ge$ 3${\farcm}$2 from the cluster center in the case
of M13 in the derivation of the observed RGB luminosity functions of M3
and M13, respectively, where $r$ is the projected distance from the
cluster center.
For the theoretical RGB luminosity function, that of Bergbusch \& VandenBerg
(2001) with age 13.5 Gyr, [Fe/H] = $-$1.54, [$\alpha$/Fe] = $+$0.30, and
$Y$ = 0.2362, which was generated from the isochrone by Bergbusch \&
VandenBerg (2001) with $B$$V$$R$$I$ color-$T_{\rm eff}$ relations as described
by VandenBerg \& Clem (2003), was adopted in both the cases of M3 and M13.

Distance moduli of M3 and M13 were derived using ZAHB
levels ($V_{\rm ZAHB}$) of M3 and M13 as a distance indicator.
The procedures for deriving the distance moduli of M3 and M13 are as
follows. Since ($m-M$)$_{V}$ equals ($V_{\rm ZAHB}$ $-$
$M_{V}^{\rm ZAHB}$), we derived $V_{\rm ZAHB}$ and $M_{V}^{\rm ZAHB}$
using equation (2) of Ferraro et al. (1999) for $V_{\rm ZAHB}$ and
equation (4) of Ferraro et al. (1999) for $M_{V}^{\rm ZAHB}$.
According to equation (2) of Ferraro et al. (1999) $V_{\rm ZAHB}$ is

\begin{equation}
V_{\rm ZAHB} = {\langle}V_{\rm HB}{\rangle} + 0.106{\rm [M/H]}^{2} +
0.236{\rm [M/H]} + 0.193~ .
\end{equation}

Also, according to equation (4) of Ferraro et al. (1999) $M_{V}^{\rm ZAHB}$ is

\begin{equation}
M_{V}^{\rm ZAHB} = 1.0005 + 0.3485{\rm [Fe/H]}_{\rm CG97} +
0.0458{\rm [Fe/H]}_{\rm CG97}^{2}~ .
\end{equation}

For ${\langle}V_{\rm HB}{\rangle}$ for M3 and M13 we used the values
derived in $\S$ 5.
${\langle}V_{\rm HB}{\rangle}$ for M3 is 15.72 ${\pm}$ 0.07 mag and the value for
M13 is 15.02 ${\pm}$ 0.10 mag. For [M/H] (${\alpha}$-element enhanced
global metallicity) and [Fe/H]$_{\rm CG97}$ ([Fe/H] in the Carretta \& Gratton
1997 scale) for
M3 and M13 we used the values listed in Ferraro et al. (1999). Since the
metallicities of M3 and M13 are nearly the same we took [Fe/H]$_{\rm CG97}$
= $-$1.39 and [M/H] = $-$1.18, respectively, for M3 and M13.
So, ($m-M$)$_{V}$ of M3 is 15.18 ${\pm}$ 0.21 mag and that of
M13 is 14.48 ${\pm}$ 0.22 mag.
These values nearly coincide with mean apparent
distance moduli derived by the subdwarf fitting technique using field
subdwarfs having $Hipparcos$ parallax measurements and with apparent
distance moduli obtained by Harris (1996) based on the mean HB luminosity
levels of M3 and M13 within the errors. According to Rood et al. (1999) the
apparent distance modulus of M3 derived by the subdwarf fitting technique with
field subdwarfs having $Hipparcos$ parallax measurements is ($m-M$)$_{V}$
= 15.19 mag, which value is adopted from private communication with R. G.
Gratton in 1998 adopting Ferraro et al.'s (1997a) photometry. According
to Gratton et al. (1997) the apparent distance modulus of M13 derived
by the subdwarf fitting technique with field subdwarfs having $Hipparcos$
parallax measurements is ($m-M$)$_{V}$ = 14.45 mag adopting Richer \&
Fahlman's (1986) photometry for the main sequence fiducial sequences.
And according to the latest database of Harris (1996), the apparent
distance moduli of M3 and M13, which are based on the mean HB luminosity
levels of M3 and M13, are ($m-M$)$_{V}$ = 15.12 mag and 14.48 mag,
respectively. The latest database of Harris (1996) adopted Johnson \&
Bolte's (1998) photometry of M3 and Paltrinieri et al.'s (1998) photometry
of M13 for the determinations of the mean HB luminosity levels of
M3 and M13, respectively. The distance moduli of M3 and M13
derived in this study and related parameters are briefly summarized in
Table 5.

\placetable{tbl-5}

From these distance moduli we shifted the  theoretical RGB luminosity
function of Bergbusch \& VandenBerg (2001) ${\Delta}V$ = 15.18 mag
for M3 and ${\Delta}V$ = 14.48 mag for M13. Normalization of
the observed RGB luminosity function of M3 and the theoretical RGB luminosity
function of Bergbusch \& VandenBerg (2001) was performed for the
$V$ =16.40--17.40 mag interval and that of
M13 was carried out for the $V$ = 15.95--16.95 mag interval.
Also, we combined these results into one figure, shifting the whole dataset
of M3 ${\Delta}V$ = $-$0.70 mag and ${\Delta}$(log $N$) = $-$0.16959
in order to make the theoretical RGB luminosity functions of
Bergbusch \& VandenBerg (2001) normalized for different clusters
coincide exactly while keeping the observed RGB luminosity function of M13 in
its absolute scale. The results are shown in Figure 7 for M3 and M13.

\begin{figure}
\figurenum{7} \epsscale{0.8} \plotone{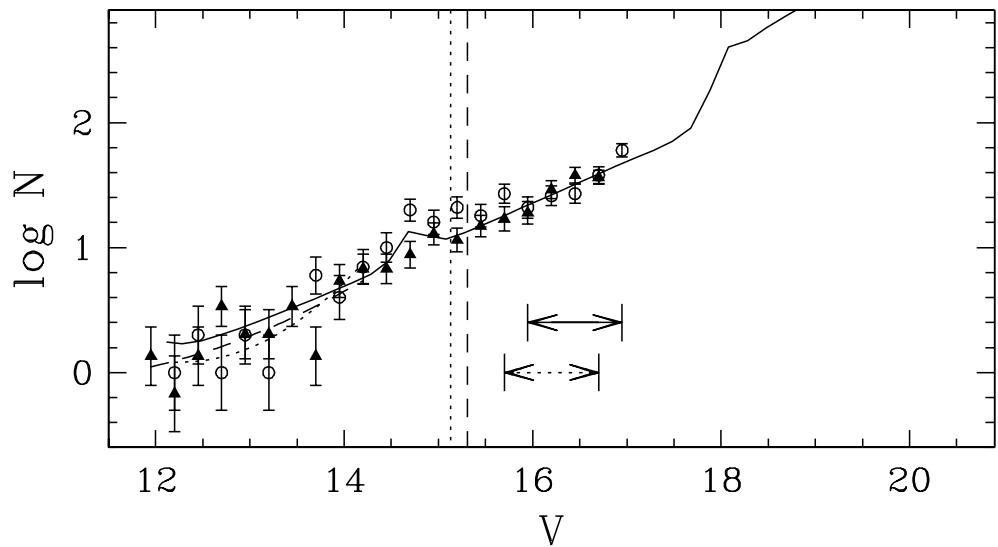}
\caption{Comparison of the observed RGB luminosity functions of M3 and M13
and the theoretical RGB luminosity function of Bergbusch \& VandenBerg (2001).
Closed triangles with error bars denote the observed RGB luminosity function
of M3, open circles with error bars denote the observed RGB luminosity
function of M13, and the solid line is the theoretical RGB luminosity
function of Bergbusch \& VandenBerg (2001). Connected dotted arrows between
two short vertical lines indicate normalization interval of the observed
RGB luminosity function of M3 and the theoretical RGB luminosity function
of Bergbusch \& VandenBerg (2001), and connected solid arrows between
two short vertical lines indicate the normalization interval of the observed
RGB luminosity function of M13 and the theoretical RGB luminosity
function of Bergbusch \& VandenBerg (2001).
Vertical dotted line indicates RGB magnitude limit when ${\Delta}BC$ = 0.11
mag ($V$ = 15.13 mag) and vertical dashed line indicates RGB magnitude
limit when ${\Delta}BC$ = 0.29 mag ($V$ = 15.31 mag). The dashed line is
the second-order least-square fitting to the observed RGB luminosity
function of M3 in the $V$ = 11.95--14.20 mag interval and the dotted line is the
second-order least-square fitting to the observed RGB luminosity function
of M13 in the $V$ = 12.20--14.20 mag interval. See text for more details.}
\label{fig7}
\end{figure}

\placefigure{fig7}

In Figure 7 closed triangles with error bars denote the observed RGB
luminosity function of M3 with projected distance from the cluster
$r$ $\ge$ 2${\farcm}$6 and the solid line is the theoretical RGB luminosity
function of Bergbusch \& VandenBerg (2001) and the error bars are 1 $\sigma$ of
Poisson statistical errors. Connected dotted arrows between two short vertical
lines indicate the normalization interval of the observed RGB luminosity
function of M3 and the theoretical RGB luminosity function of
Bergbusch \& VandenBerg (2001). The dashed line is the second-order
least-square fitting to the observed RGB luminosity function of M3 in the
$V$ = 11.95--14.20 mag interval (originally in the $V$ = 12.65--14.90 mag
interval when unshifted). Although according to the second-order
least-square fitting line, M3 shows a slight deficiency of stars relative
to the theoretical RGB luminosity function of Bergbusch \& VandenBerg (2001)
in the bright region, the number of deficient stars is only 1 $\pm$ 6 in its
absolute sense, where the error arises from the Poisson statistical error
of the observed RGB luminosity function of M3.
However, in general the observed RGB luminosity function of M3 and the
theoretical RGB luminosity function of Bergbusch \& VandenBerg (2001)
seem to agree quite well over the whole magnitude range and there are
no extra stars in the observed RGB luminosity function of M3 relative to
the theoretical RGB luminosity function of Bergbusch \& VandenBerg (2001).
According to Figure 7, the total number of stars in the observed RGB luminosity
function of M3 in the $V$ = 11.95--16.70 mag range (originally in the
$V$ = 12.65--17.40 mag range when unshifted) is 328 $\pm$ 18 in its
absolute sense and the
total number of stars in the theoretical RGB luminosity function of
Bergbusch \& VandenBerg (2001) in the same magnitude range is estimated
to be 332 in its absolute sense. So, their difference is only 4
(1$\%$ of the total number of stars
in the observed RGB luminosity function of M3) and they are in quite
excellent agreement to within a 99$\%$ confidence level with the total number
of stars of the observed RGB luminosity function of M3.

In Figure 7, open circles with error bars denote the observed RGB luminosity
function of M13 with the projected distance from the cluster center
$r$ $\ge$ 3${\farcm}$2 and the error bars represent 1 $\sigma$ Poisson
statistical errors. Connected arrows between two short vertical lines
indicate the normalization interval of the observed RGB luminosity function for
M13 and the theoretical RGB luminosity function of Bergbusch \& VandenBerg
(2001). According to Figure 7, the observed luminosity function of M13 in
the bright region shows a slight deficiency of stars relative to the
theoretical RGB luminosity function of Bergbusch \& VandenBerg (2001).
The dotted line is the second-order least-square fitting to the observed
RGB luminosity function of M13 in the $V$ = 12.20--14.20 mag interval.
The second-order least-square fitting line of M13 also shows a slight
deficiency of stars relative to the theoretical RGB luminosity function
of Bergbusch \& VandenBerg (2001) in the bright region and the number of
deficient stars of the observed RGB luminosity function is 4 $\pm$ 5.
According to Figure 7, the observed RGB luminosity function of M13 shows
``extra stars'' in the $V$ = 14.70--15.70 mag range relative to the
theoretical RGB luminosity function of Bergbusch \& VandenBerg (2001).
The number of extra stars is estimated to be 31 $\pm$ 10 with about
3 $\sigma$ Poisson statistical error. According to Figure 7,
``extra stars'' are partially included in the magnitude range where RGB
stars are used to derive the population ratio $R$. The number of extra
stars included in the RGB stars used in the derivation of population ratio
$R$ is estimated to be 17 $\pm$ 7 when ${\Delta}BC$ = 0.11 mag in the
$V$ = 14.70--15.13 mag range, and 20 $\pm$ 8 when ${\Delta}BC$ = 0.29 mag
in the $V$ = 14.70--15.31 mag range. According to Figure 7, the total number
of stars in the observed RGB luminosity function of M13 in the $V$
= 12.20--16.95 mag range is 308 $\pm$ 18 and total number of stars in the
theoretical RGB luminosity function of Bergbusch \& VandenBerg (2001)
in the same magnitude range is estimated to be 273. So, their difference
is 35, or 11$\%$ of the total number of stars in the observed RGB luminosity
function of M13 and they do not agree within Poisson statistical error.
This fact also supports the contention that there are ``extra stars'' in
the observed RGB luminosity function of M13.

In order to determine whether the ``extra stars'' in the observed RGB
luminosity function
of M13 relative to the theoretical RGB luminosity function of Bergbusch
\& VandenBerg (2001) are real even when we compare with the observed
RGB luminosity function of M3 in a differential sense, we intercompare
the observed RGB luminosity functions of M3 and M13.
According to Figure 7, the number of deficient stars in the
observed RGB luminosity function of M13 with respect to the observed RGB
luminosity function of M3 in the $V$ = 12.20--14.20 mag range is 2 $\pm$ 11
and is therefore negligible. In the $V$ = 14.45--15.70 mag range the observed
RGB luminosity function of M13 shows ``extra stars'' with respect to the
observed RGB luminosity function of M3. The number of extra stars in M13 is
estimated to be 40 $\pm$ 21 in the following ways.
In the $V$ = 14.45--15.70 mag range there are six bins for the
RGB luminosity functions of M3 and M13. So, ``extra stars'' in M13 are
calculated in each bin and their associated errors are also calculated in
each bin according to the standard error propagation law with
Poisson statistical errors for each bin of M3 and M13. In each
bin ``extra stars'' of M13 are calculated simply as the number of M13 stars
in the RGB luminosity function of M13 minus the number of M3 stars in the
RGB luminosity function of M3 normalized to the M13 plane. And finally the
``extra stars'' of M13 in six bins are simply added and their associated
errors in six bins are simply combined according to the standard error
propagation law. Although it seems that Poisson statistical
error overestimates the counting error, the number of extra stars in M13 is
still at the 2 $\sigma$ level of Poisson statistical
error and about 13$\%$ of the total number of stars of the observed RGB
luminosity function of M13. According to Figure 7, the ``extra stars'' of
M13 are partially included in the magnitude range where RGB stars are
used in the derivation of population ratio $R$. The number of extra stars
included in the RGB stars used in the derivation of the population ratio $R$
is estimated to be 24 $\pm$ 15 when ${\Delta}BC$ = 0.11 mag in the
$V$ = 14.45--15.13 mag range, and 28 $\pm$ 17 when ${\Delta}BC$ = 0.29 mag
in the $V$ = 14.45--15.31 mag range.

Because in the case of M13 we estimated the number of the extra stars
of the RGB of M13 included in the calculation of $R$ ratios in the
comparison of the observed RGB luminosity functions of M3 and M13, we
can estimate the $R$ ratios for M13 excluding the effects of the extra
stars of RGB. The $R$ ratios for M13
excluding the effects of the extra stars of the RGB of M13 are estimated
as follows. In the case where only the normal HB stars are included, $R$ for M13
is 57/(74$_{\rm -5}^{\rm +10}-(24 \pm 15)$) = 57/(50$_{\rm -16}^{\rm +18}$)
= 1.140$_{\rm -0.302}^{\rm +0.536}$ when ${\Delta}BC$ = 0.11 mag
and 57/(92$_{\rm -11}^{\rm +4}-(28 \pm 17)$) = 57/(64$_{\rm -20}^{\rm +17}$)
= 0.891$_{\rm -0.187}^{\rm +0.405}$ when ${\Delta}BC$ = 0.29 mag in
the $r$ $\ge$ 3${\farcm}$2 range. These $R$ ratios are smaller than the
$R$ ratios for M3. If we assume that the helium abundances of M3 and
M13 are nearly the same, including only the normal HB stars in the $R$ ratios
seems to overcorrect the prolonged lifetime of the extended BHB stars
of M13. In M13 the second BHB gap is at $V$ $\approx$ 16.80 mag. The number
of HB stars of M13 between the first BHB gap and the second BHB gap (16.00 ${<}$
$V$ ${<}$ 16.80 mag interval) is 24 in the $r$ $\ge$ 3${\farcm}$2 range. If we
add this number to the $N_{\rm HB}$ of M13, the population ratios $R$ and
$R_{\rm 2}$ for M13 are as follows. $R$ for M13 is
(57 + 24)/(74$_{\rm -5}^{\rm +10}-(24 \pm 15)$) = 81/(50$_{\rm -16}^{\rm +18}$)
= 1.620$_{\rm -0.429}^{\rm +0.762}$ when ${\Delta}BC$ = 0.11 mag and
(57 + 24)/(92$_{\rm -11}^{\rm +4}-(28 \pm 17)$) = 81/(64$_{\rm -20}^{\rm +17}$)
= 1.266$_{\rm -0.266}^{\rm +0.575}$ when ${\Delta}BC$ = 0.29 mag in the
$r$ $\ge$ 3${\farcm}$2 range. $R_{\rm 2}$ for M13 is 9/(57 + 24) = 9/81
= 0.111 in the $r$ $\ge$ 3${\farcm}$2 range. In this case these population
ratios $R$ and $R_{\rm 2}$ for M13 in the $r$ $\ge$ 3${\farcm}$2 range
are consistent with population ratios
$R$ and $R_{\rm 2}$ for M3 in the $r$ $\ge$ 2${\farcm}$6 range.
In the case where all the HB stars are included, $R$ for M13 is
127/(74$_{\rm -5}^{\rm +10}-(24 \pm 15)$) = 127/(50$_{\rm -16}^{\rm +18}$)
= 2.540$_{\rm -0.672}^{\rm +1.195}$ when ${\Delta}BC$ = 0.11 mag and
127/(92$_{\rm -11}^{\rm +4}-(28 \pm 17)$) = 127/(64$_{\rm -20}^{\rm +17}$)
= 1.984$_{\rm -0.416}^{\rm +0.902}$ when ${\Delta}BC$ = 0.29 mag in the
$r$ $\ge$ 3${\farcm}$2 range. These $R$ ratios for M13 are about
0.90 units higher than those for M3 although error levels are
large. This value clearly imply that in GCs having the extended BHB stars
we must correct anyway the prolonged lifetime of hot HB stars in $R$
ratios to estimate the true helium abundances of these kinds of GCs.
If we assume that the helium abundances of M3 and M13 are nearly the
same, $\approx$0.90 units are larger than the $\approx$0.25 units which
are expected for the increased values of the $R$ ratios for GCs which have
blue HB morphologies according to Zoccali et al. (2000). This fact
may imply that the actual lifetime of the extended BHB stars is larger
than that predicted by the theoretical hot HB stellar models
(Castellani et al. 1994; Zoccali et al. 2000). Otherwise, the helium
abundance of M13 must be larger than that of M3.

\section{Conclusion and Summary}

We presented $V$ versus $B$$-$$V$, $V$ versus $V$$-$$I$, and $V$ versus
$B$$-$$I$ CMDs for M3 and M13. In these CMDs the AGB stars of M3 and M13 are
the most clearly separated from RGB stars in $V$ versus $B$$-$$I$ CMDs.
We have found stars widely distributed over the RR Lyrae instability strip,
and also some stars below the level of ZAHB in both the RHB and BHB of
M3. When 84 known
RR Lyraes are identified from the observed field according to the
M3 variable stars list, these stars are also identified as RR Lyraes in
the M3 CMDs. So, these stars are also RR Lyrae variable stars in their
close minimum or minimum pulsational stages in light of their $V$ magnitude.
In M13 we have found a few stars at the RGB tip,
which tend slightly to the redder and fainter direction
and BHB stars are well populated down to the MSTO with at least two
gaps. We have found AGB bumps at $V$$_{\rm AGB}^{\rm bump}$ = 14.85 ${\pm}$
0.05 mag for M3 and at $V$$_{\rm AGB}^{\rm bump}$ = 14.25 ${\pm}$ 0.05
mag for M13. AGB stars in M3 are more concentrated
near the AGB bump, while those in M13 are scattered along the AGB sequence.
We identified RGB bumps for M3 and M13 whose positions are
$V$$_{\rm RGB}^{\rm bump}$ = 15.50 $\pm$ 0.05 mag for M3 and
$V$$_{\rm RGB}^{\rm bump}$ = 14.80 $\pm$ 0.05 mag for M13 from the slope
changes of the integrated luminosity functions of M3 and M13.

We presented population ratios $R$ (= $N_{\rm HB}$/$N_{\rm RGB}$),
and $R_{\rm 2}$ (= $N_{\rm AGB}$/$N_{\rm HB}$) for M3 and M13 using
their $V$ versus $B$$-$$I$ CMDs in Figures 1{\it c} and 2{\it c} for
M3 and M13, respectively. $R$ for M3 is 1.414$_{\rm -0.016}^{\rm +0.086}$ or
1.281$_{\rm -0.121}^{\rm +0.056}$ in the $r$ $\ge$ 2${\farcm}$6 range
depending on the adopted differential bolometric corrections (${\Delta}BC$)
between ${\langle}V_{\rm HB}{\rangle}$ at log $T_{\rm eff}$ = 3.85
and RGB at the ${\langle}V_{\rm HB}{\rangle}$ level. $R_{\rm 2}$ for
M3 is 0.122 in the $r$ $\ge$ 2${\farcm}$6 range.
$R$ for M13 is 0.770$_{\rm -0.091}^{\rm +0.056}$ or
0.620$_{\rm -0.026}^{\rm +0.084}$ in the $r$ $\ge$ 3${\farcm}$2 range
depending on the adopted differential bolometric corrections (${\Delta}BC$)
between ${\langle}V_{\rm HB}{\rangle}$ at log $T_{\rm eff}$ = 3.85
and RGB at the ${\langle}V_{\rm HB}{\rangle}$ level when only
normal HB stars are included in $N_{\rm HB}$. $R_{\rm 2}$ for M13 is 0.158 
in the $r$ $\ge$ 3${\farcm}$2 range
when only normal HB stars are included in $N_{\rm HB}$.
In the case where all the HB stars are included in $N_{\rm HB}$ for
the population ratios, $R$ for M13 is 1.716$_{\rm -0.204}^{\rm +0.125}$
or 1.380$_{\rm -0.057}^{\rm +0.188}$ in the $r$ $\ge$ 3${\farcm}$2 range
depending on the adopted differential bolometric corrections (${\Delta}BC$)
between ${\langle}V_{\rm HB}{\rangle}$ at log $T_{\rm eff}$ = 3.85
and the RGB at the ${\langle}V_{\rm HB}{\rangle}$ level. In this case
$R_{\rm 2}$ for M13 is 0.071 in the $r$ $\ge$ 3${\farcm}$2 range.

When comparing the population ratios $R$ and $R_{\rm 2}$ for M3 and M13
in the case where only normal HB stars are included in $N_{\rm HB}$
for M13, $R$ values for M3
are larger than those for M13. The small values of
$R$ for M13 compared to those for M3 could be
explained partially by the increased number of RGB stars in M13
due to an effective extra deep mixing which prolongs the RGB phase,
and partially by
the cutoff of BHB tail stars below the first BHB gap, which have lost
practically all the H-rich envelope. However $R_{\rm 2}$ values for M3 and
M13 are similar and are in the range of the average value
for the 43 Galactic GCs ${\langle}R_{\rm 2}{\rangle}$ = 0.162 $\pm$ 0.059
of Sandquist (2000), and in good agreement with the value ${\langle}R_{\rm 2}{\rangle}$
= 0.15 derived by the more recent models of Cassisi et al. (1998).
When comparing the population ratios $R$ and $R_{\rm 2}$ for M3 and M13
in the case where all the HB stars are included in $N_{\rm HB}$ for
M13, the $R$ values for M3 and M13 are similar. However, the $R_{\rm 2}$
values for M3 are about a factor of 2 larger than those for M13,
which fact can be explained by the prolonged lifetime of BHB
tail stars of M13 (Castellani et al. 1994; Zoccali et al. 2000).
Since the lifetime of BHB tail stars of M13 is prolonged, $R$ for M13
must be larger than for M3 but this contradicts
the actual $R$ values for M3 and M13. This fact could be explained
by the increased number of RGB stars in M13 due to an effective extra
deep mixing which prolongs the RGB phase rendering $R$ for M13 smaller
than the expected value.

We compared the observed RGB luminosity functions of M3 and M13 with the
theoretical RGB luminosity function of Bergbusch \& VandenBerg (2001)
for exactly the same radial distances from the cluster centers in the
derivation of population ratios $R$ and $R_{\rm 2}$. We discovered
``extra stars'' in the observed RGB luminosity function
of M13 which are expected from comparisons of population ratios
$R$ and $R_{\rm 2}$ for M3 and M13. The increased number of stars
(= ``extra stars'') in the RGB luminosity function of M13
may be the result of an effective ``deep mixing'' as suggested by
Langer, Bolte, \& Sandquist (2000). This suggests that ``deep mixing''
may be another possible second parameter responsible for the different HB
morphologies of M3 and M13. However, the ``extra stars'' due to deep mixing
may be found in the upper RGB rather than the RGB bump. This study found 
extra stars near the RGB bump only.

In the case of M13 we estimated the number of the extra stars of the RGB
of M13 included in the calculation of $R$ ratios in the comparison of the
observed RGB luminosity functions of M3 and M13. So, we can estimate
the $R$ ratios for M13 excluding the effects of the extra stars of RGB.
When we exclude the effects of the extra stars of the RGB of M13 and assume
that the helium abundances of M3 and M13 are the same, $R$ and $R_{\rm 2}$
for M3 and M13 in the considered radial distance from the cluster centers
respectively become similar if we include only the HB stars brighter
than the second BHB gap ($V$ $\approx$ 16.80 mag) of M13 in $N_{\rm HB}$
of M13.

Recently, Sneden et al. (2004) argued that the severe abundance anomaly
of oxygen for bright RGB stars and the excessive blueness of the HB of M13 are
not due to deep mixing and that the severe abundance anomaly of oxygen
for the bright RGB stars of M13 is due to pollution by the nuclear processed
material ejected from cluster stars in the 3--6 $M_{\sun}$ range.
Although [Fe/H] of M3, M13, and NGC 6752 are similar and
the HB morphologies of M13 and NGC 6752 are similar in that they have no
RHB stars, few or no RR Lyrae stars and predominantly BHB stars
extending down to $\sim$4 mag below the HB level while M3 has a
uniformly populated HB morphology, the abundance patterns are similar
between M3 and NGC 6752 rather than between M13 and NGC 6752. Moreover,
NGC 6752 shows abundance anomalies already in the main sequence stars
and M13 and NGC 6752 satisfy the same anticorrelation of [O/Fe] with the
magnesium isotopic ratio, which could only have come about in nuclear
processed material ejected from cluster stars in the 3--6 $M_{\sun}$ range.

According to Sweigart (1997), ``deep mixing'' could increase the
envelope abundance, and helium mixing increases the RGB tip luminosity
and leads to enhanced mass loss along the RGB. These effects
largely influence the subsequent horizontal branch evolution.
They can produce a bluer HB morphology, making it possible to explain the hot
HB population found in M13 as well as the difference in HB morphology
between M3 and M13.
Although the subsequent work of Caloi (2001) and the spectroscopic
study on hot stars in M3 and M13 by Moehler et al. (2003) could
not confirm that the ``helium mixing'' scenario of Sweigart (1997) is
a definite explanation for the origin of the blue HB tail in M13, it seems
worth reexamining helium mixing as a partial cause for the
morphological difference in the normal HB between M3 and M13.

We conclude that the second parameter pair, M3 and M13, differ not
only in their HB morphologies but also in their AGB sequence characteristics and
RGB luminosity functions. It is quite probable that the cause of
extra stars in the RGB luminosity function of M13, which may be due to
deep mixing, may lead to differences in subsequent sequences from
those of M3.

\acknowledgments
This work was financially supported by the Korean Astronomy Observatory.
This work was partially supported by grant No. R04-2002-000-00138-0 from
the Basic Research Program of the Korea Science \& Engineering Foundation.

\clearpage

\clearpage

\clearpage

\clearpage

\clearpage

\clearpage

\clearpage

\begin{deluxetable}{lcrrcc} 
\tablenum{1} \tablecaption{Observations Summary of M3 and M13.
\label{tbl-1}} \tablewidth{0pt}
\tablehead{\colhead{} & \colhead{} & \colhead{Exposure Time} &
\colhead{Seeing (FWHM)} & \colhead{} & \colhead{} \\
\colhead{Object} & \colhead{Filter} & \colhead{(s)} &
\colhead{(arcsec)} & \colhead{Air Mass Range} &
\colhead{Observing Date} }
\startdata M3 &  $B$  & 2 ${\times}$ 160/2 ${\times}$ 800 &
2.2/2.0  & 1.01 ${\sim}$ 1.01 & 2001 May 10 \\
  &  $V$ & 2 ${\times}$ 120/2 ${\times}$ 600 &
2.0/2.0  & 1.02 ${\sim}$ 1.04 & 2001 May 10 \\
  &  $I$  & 2 ${\times}$ 20/2 ${\times}$ 60/2 ${\times}$ 300 &
1.8/1.9/1.9     & 1.05 ${\sim}$ 1.10 & 2001 May 10 \\
  M13 & $B$  & 2 ${\times}$ 160/2 ${\times}$ 800 &
1.9/1.8  & 1.03 ${\sim}$ 1.06 & 2001 May 11 \\
      & $V$ &  2 ${\times}$ 20/2 ${\times}$ 120 &
1.8/1.8  & 1.13 ${\sim}$ 1.15 & 2001 May 11 \\
      & $I$  & 2 ${\times}$ 20/2 ${\times}$ 60 &
1.4/1.6  & 1.22 ${\sim}$ 1.25 & 2001 May 11 \\
\enddata

\end{deluxetable}

\begin{deluxetable}{cccccc}
\tablenum{2} \tablecolumns{6} \tablecaption{Population Ratios of M3.
\label{tbl-2}} \tablewidth{0pt}
\tablehead{\colhead{${\langle}V_{\rm HB}{\rangle}$} & \colhead{$N_{\rm HB}$} &
\colhead{$N_{\rm AGB}$} & \colhead{$N_{\rm RGB}$} &
\colhead{$R$ (= $N_{\rm HB}$/$N_{\rm RGB}$)} &
\colhead{$R_{\rm 2}$ (= $N_{\rm AGB}$/$N_{\rm HB}$)}}
\startdata
\multicolumn{6}{c}{($r$ ${\ge}$ 2${\farcm}$6)} \\
\multicolumn{6}{c}{${\Delta}BC$ = 0.11 mag case (Ferraro et al. 1997a)} \\
\tableline
15.72 ${\pm}$ 0.07 & 123 & 15 & 87$_{\rm -5}^{\rm +1}$ &
 1.414$_{\rm -0.016}^{\rm +0.086}$ & 0.122 \\
\tableline
\multicolumn{6}{c}{${\Delta}BC$ = 0.29 mag case (Sandquist 2000)} \\
\tableline
15.72 ${\pm}$ 0.07 & 123 & 15 & 96$_{\rm -4}^{\rm +10}$ &
 1.281$_{\rm -0.121}^{\rm +0.056}$ & 0.122 \\
\enddata
\end{deluxetable}

\begin{deluxetable}{cccccc}
\tablenum{3} \tablecolumns{6} \tablecaption{Population Ratios of M13 (HB Stars
above first Gap only).
\label{tbl-3}} \tablewidth{0pt}
\tablehead{\colhead{${\langle}V_{\rm HB}{\rangle}$} & \colhead{$N_{\rm HB}$} &
\colhead{$N_{\rm AGB}$} & \colhead{$N_{\rm RGB}$} &
\colhead{$R$ (= $N_{\rm HB}$/$N_{\rm RGB}$)} &
\colhead{$R_{\rm 2}$ (= $N_{\rm AGB}$/$N_{\rm HB}$)}}
\startdata
\multicolumn{6}{c}{($r$ ${\ge}$ 3${\farcm}$2)} \\
\multicolumn{6}{c}{${\Delta}BC$ = 0.11 mag case (Ferraro et al. 1997a)} \\
\tableline
15.02 ${\pm}$ 0.10 & 57 & 9 & 74$_{\rm -5}^{\rm +10}$ &
 0.770$_{\rm -0.091}^{\rm +0.056}$ & 0.158 \\
\tableline
\multicolumn{6}{c}{${\Delta}BC$ = 0.29 mag case (Sandquist 2000)} \\
\tableline
15.02 ${\pm}$ 0.10 & 57 & 9 & 92$_{\rm -11}^{\rm +4}$ &
 0.620$_{\rm -0.026}^{\rm +0.084}$ & 0.158 \\
\enddata
\end{deluxetable}

\begin{deluxetable}{cccccc}
\tablenum{4} \tablecolumns{6} \tablecaption{Population Ratios of M13 (all HB
Stars).
\label{tbl-4}} \tablewidth{0pt}
\tablehead{\colhead{${\langle}V_{\rm HB}{\rangle}$} & \colhead{$N_{\rm HB}$} &
\colhead{$N_{\rm AGB}$} & \colhead{$N_{\rm RGB}$} &
\colhead{$R$ (= $N_{\rm HB}$/$N_{\rm RGB}$)} &
\colhead{$R_{\rm 2}$ (= $N_{\rm AGB}$/$N_{\rm HB}$)}}
\startdata
\multicolumn{6}{c}{($r$ ${\ge}$ 3${\farcm}$2)} \\
\multicolumn{6}{c}{${\Delta}BC$ = 0.11 mag case (Ferraro et al. 1997a)} \\
\tableline
15.02 ${\pm}$ 0.10 & 127 & 9 & 74$_{\rm -5}^{\rm +10}$ &
 1.716$_{\rm -0.204}^{\rm +0.125}$ & 0.071 \\
\tableline
\multicolumn{6}{c}{${\Delta}BC$ = 0.29 mag case (Sandquist 2000)} \\
\tableline
15.02 ${\pm}$ 0.10 & 127 & 9 & 92$_{\rm -11}^{\rm +4}$ &
 1.380$_{\rm -0.057}^{\rm +0.188}$ & 0.071 \\
\enddata
\end{deluxetable}

\begin{deluxetable}{rcccccc}
\tablenum{5} \tablecaption{Distance Moduli of M3 and M13 Derived in
This Study and Related Parameters.
\label{tbl-5}} \tablewidth{0pt}
\tablehead{\colhead{Cluster} & \colhead{[Fe/H]$_{\rm CG97}$} & \colhead{[M/H]}
 &  \colhead{${\langle}V_{\rm HB}{\rangle}$} &  \colhead{$V_{\rm ZAHB}$}
 &  \colhead{$M_{V}^{\rm ZAHB}$}  &   \colhead{($m-M$)$_{V}$}}
\startdata
  M3 &  $-$1.39  &  $-$1.18 & 15.72 $\pm$ 0.07 & 15.78 $\pm$ 0.07 &
  0.60 &  15.18 $\pm$ 0.21  \\
 M13 &  $-$1.39  &  $-$1.18 & 15.02 $\pm$ 0.10 & 15.08 $\pm$ 0.10 &
  0.60 &  14.48 $\pm$ 0.22  \\
\enddata
\end{deluxetable}

\end{document}